# The empirical study of e-learning post-acceptance after the spread of COVID-19: A multi-analytical approach based hybrid SEM-ANN

**Ashraf Elnagar**[1,4]**, Imad Afyouni**[1,4]**, Ismail Shahin**[2,4]**, Ali Bou Nassif**[3,4] **and Said A. Salloum**[4,5*]

[1]Computer Science Department, University of Sharjah, Sharjah, UAE
[2]Electrical Engineering Department, University of Sharjah, Sharjah, UAE
[3]Computer Engineering Department, University of Sharjah, Sharjah, UAE
[4]Machine Learning and NLP Research Group, University of Sharjah, Sharjah, UAE
[5] School of Computing & Science & Engineering, University of Salford, UK

{ashraf, iafyouni, ismail, anassif, ssalloum}@sharjah.ac.ae
*Corresponding Author: Said A. Salloum

**Abstract**

There are several reasons why the fear of vaccination has caused population-rejection. Questions have been raised by students regarding the effectiveness of vaccines, which in turn has led to vaccination hesitancy. Students' perceptions are influenced by vaccination hesitancy, which affects the acceptance of e-learning platforms. Hence, this research aimed to examine the post-acceptance of e-learning platforms on the basis of a conceptual model that employs different variables. Distinct contribution is made by every variable to the post-acceptance of e-learning platforms. A hybrid model was used in the current study in which technology acceptance model (TAM) determinants were employed along with other external factors such as "fear of vaccination, perceived routine use, perceived enjoyment, perceived critical mass, and self-efficiency" which are directly linked to "post-acceptance of e-learning platforms". The focus of earlier studies on this topic has been on the significance of e-learning acceptance in various environments and countries. However, in this study, the newly-spread use of e-learning platforms in the gulf area was examined using a hybrid conceptual model. The empirical studies carried out in the past mainly used structural equation modelling (SEM) analysis; however, this study used an evolving hybrid analysis approach, in which SEM and the artificial neural network (ANN) that are based on deep learning were employed. The importance-performance map analysis (IPMA) was also used in this study to determine the significance and performance of each factor. The proposed model is backed by the findings of data analysis. It is shown in the findings that "fear of vaccination, perceived ease of use, perceived usefulness, perceived routine use, perceived enjoyment, perceived critical mass, and self-efficiency" significantly affect students' behavioral intention to utilise e-learning platforms for educational objectives. It is also shown in the analysis of ANN as well as IPMA that perceived ease of use is the most significant predictor of post-acceptance of e-learning platforms. Theoretically, sufficient explanations have been offered by the suggested model regarding the factors that influence the post-acceptance of e-learning platforms in terms of the internet service factors at the individual level. In the practical sense, these findings would help decision-makers and practitioners in higher educational institutions identify those factors that should be given more significance compared to others and plan their policies appropriately. Methodologically, the ability of the deep ANN architecture to identify the non-linear relationships between the factors involved in the theoretical model has been determined in this research. The implication offers extensive information about taking effective steps to decrease the fear of vaccination among people and increase vaccination confidence among teachers, educators, and students, which will consequently have an impact on the entire society.

1. Introduction

Various challenges were experienced by higher education institutions during the COVID-19 pandemic, even after the availability of vaccines. Due to these challenges, a major change was required in the field of teaching and learning. During the pandemic, majority of the schools and universities shifted to virtual classrooms from the traditional physical classroom, which was the only way teaching methods and goals could be applied [1]–[3]. University students shifted to an entirely different setting, where they experienced a number of issues [4], [5]. When the challenges experienced by students are comprehended, the method that can be used to evaluate the students' understanding, success and accomplishment during the second wave of the coronavirus pandemic can be established [6].

It should be noted that this challenge continues to be experienced even after the vaccination has become available. This arises from the fact that the population has rejected the vaccination. Vaccination hesitancy refers to the hesitation shown by people towards different vaccines in several countries worldwide. It is a critical factor that impacts the world in general and the education system in specific. It suggests that a delay is experienced in the acceptance or rejection of vaccines, even though vaccines have become available. This view differs across certain locations and over time and is closely linked to confidence and convenience. It is also influenced by contextual factors, group impact, individual beliefs, and people's faith in health services. Vaccination hesitancy differs from one specific user to another, based on their self-perception, perceived risk, and workplace. Therefore, vaccination hesitancy is the key issue being experienced regarding the COVID-19 vaccine uptake, which means that several hurdles will be experienced by the vaccine against COVID-19 in a post-crisis scenario [7]–[9]. Opposing vaccination hesitancy is vaccination confidence, which refers to something that can be accomplished "in itself". Confidence in vaccines relies on the trust in the healthcare system as well as trust in a socio-political context. Users may not have a consistent confidence in vaccination because of the perceived risks associated with immunisation. Rather, it may give rise to smaller vaccination coverage and loss of immunity [10]–[12].

On the basis of the aforementioned assumptions, the objective of this research was to examine how vaccination hesitancy or fear affects post-acceptance of e-learning platforms, where the challenge continues to be critical and obvious because of vaccination rejection. A conceptual framework was formulated with the goal of attaining the study objective, which caters to the two factors being discussed, "i.e. post-acceptance and fear of vaccination". A few external variables were included in the conceptual model, because they have a direct relationship with post-acceptance, namely "perceived enjoyment, perceived routine use, self-efficiency and fear of vaccinations" [13]–[16]. Consequently, the precise contribution of the present study can be summed up as follows: the study first examined the impact of vaccination fear or hesitancy in the educational sector. For this, an integrated research model was used; a conceptual model was formulated in which the TAM acceptance model [17] was integrated with the flow theory [18], [19] to show the importance and predictability of the findings. Second, in contrast to the earlier empirical studies that mainly depended on structural equation modeling (SEM) analysis, an emerging hybrid analysis approach was used in this method that employs SEM and artificial neural network (ANN) based on deep learning. The importance-performance map analysis (IPMA) was also used in this research to determine the significance and performance of every factor. Third, vaccination fear and hesitancy is a critical factor and has different effects based on aspects such as the age and gender of people and the financial status of a country. A recent study revealed that there is greater vaccine hesitancy in low-income countries and among young females and older adults [20]. Lastly, the model was extended in this study to consider external variables that have a close relationship with the post-acceptance stage [20], which include critical mass and daily routine. We believe that this is the first study that seeks to examine the post-acceptance of e-learning platforms based on an integrated model in which fear of vaccination is the key variable, with the aim of filling a major research gap in the relevant literature.

## 2. Related Literature

According to the literature review, the impact of COVID-19 on the various educational e-learning platforms have been analysed in earlier studies. These platforms included Moodle, Zoom Microsoft Teams, Google Classroom, virtual reality applications, and so on. The studies showed that all these platforms were effective while the pandemic was spreading and offered a viable solution to the issue [21]–[23]. In most of the previous studies, TAM has been the most influential model used. The focus of majority of the studies was on two of the most influential constructs, i.e. perceived usefulness and perceived ease of use. The studies examined the effective part played by the two constructs in ensuring that the students' adoption or acceptance is on-demand [22], [24]–[26]. UTAUT, which is an extended model of TAM, has also been employed as a model to determine how effective the constructs have been during the pandemic. Different technology acceptance models were used in the studies carried out in India. However, the model was extended by [21] by including SUS, which is vital for examining perceived usability. The same trend was followed in the study by [23], in which a few external factors were included to extend the TAM model, such as innovativeness, computer self-efficacy, computer anxiety, social norms, perceived enjoyment, content and system quality.

The impact of the pandemic has extended to various parts of the world, which is why the study is varied in its location. Different studies were carried out in Indonesia, China, Vietnam, and Malaysia. In all of these studies, the impact of e-learning platforms during the pandemic was explained with the help of surveys or online questionnaires that were distributed to students of undergraduate educational institutions [24], [27]–[29]. The studies conducted in Romania and Europe also used surveys or online questionnaires distributed among either students or farmers.

They had a distinct sample because of the distinct objectives of their study. The study in Europe had the objective of describing the farmers' readiness to use new technology during the pandemic, while the one in Romania focused on examining how the online platform affected a sample of students during the pandemic [23], [30]. The impact of COVID-19 on the educational environment of different cities in India was examined by various researchers. It was deduced in these studies that e-learning platforms were effective in maintaining direct and indirect means of communication between the various participants in educational institutions [21], [22].

With respect to the methodology, most previous studies used only the single-stage linear data analysis [31], specifically the SEM technique. It was only possible to identify the linear correlations between the factors in the theoretical model using the single stage of SEM analysis, and this was not sufficient to predict the complex decision-making processes [32]. This limitation was addressed by a few scholars by using the ANN method as a second stage of analysis [33]–[35].

However, this approach comprises only a single hidden layer and is considered as a shallow type of ANN [36]. It has been indicated that the deep ANN architecture should be used rather than the shallow ANN, as it may improve the accuracy of non-linear models by employing over a single hidden layer [37]. Therefore, the present research uses a hybrid SEM-ANN technique that is based on a deep ANN approach which offers deep learning. The previous studies [21], [22], [38]–[40], [23]–[30] have clearly shown that majority of them were carried out in educational institutions where the teaching and learning process is mainly carried out through the e-learning platforms. In this way, it ensures that the shift from the traditional classroom to the e-learning environment occurs safely and effectively [2], [3]. In the end, this would enable all the educational institutions to accomplish their aims and objectives [26], [27].

## 3. The Development Model
### 3.1 Perceived Vaccination Fear (FV)

Within various populations, the COVID-19 vaccination fear and hesitancy is quite high, mainly due to the increase in the coronavirus conspiracy theory. Hence, the vaccination is rejected, and there is a constant increase in the vaccination hesitancy percentage [41]. Generally, the vaccination acceptance is influenced by the risk theory. There are risk regulation and cultural assessments associated with the risk theory. The risk response is quite deep considering the emotions present due to cultural developments. Risk information is evaluated in a way that the expected utility is maximised [42]–[45]. The perceived vaccination fear differs according to different genders. For women, generally, the vaccine is associated with negative experiences regarding the healthcare institution; however, the men's attitude is associated with the immune systems; they believe that their immune systems would be weakened [46], [47]. Regarding the COVID-19 vaccine, it is observed that vaccine rejection is significantly influenced by the health literacy. Research indicates that the vaccine may be rejected by the students, specifically female students, considering their health literacy. Health-protective behaviour is adopted, since their fear towards the COVID-19 vaccine is quite high. The perceived vaccination fear is what leads to the spread of the COVID-19 infection [48]–[50]. Thus, considering the earlier assessments, the vaccination fear hypothesis was developed.

**H1:** FV has a positive and significant impact on POA.

### 3.2 TAM Theory

The technology acceptance model (TAM) theory was formulated by Fred Davis, contributing to the idea of technology adoption, acceptance and post-acceptance. Perceived ease of use and perceived usefulness are the constructs of this model, and these are taken to be conceptual factors that contribute to the post-acceptance of the technology. The perceived ease of use variable is linked to the effectiveness of the easiness factor on the users' performance, while perceived usefulness is concerned with the idea of "effort-free" that improves the users' performance [51]. This led to the following hypotheses:

**H2:** PU has a positive and significant impact on POA.

**H3:** PE has a positive and significant impact on POA.

### 3.3 Perceived Daily Routine (PR)

The idea of daily routine in this context refers to the degree to which technology is a part of the routine tasks and the inclusion of technology into the standard work routine of users. The daily routine use of technology is described as the use of technology such that it enters into the daily pattern and is considered as a standard element in the life of users [13], [52], [53]. The daily routine is a significant factor that affects the post-acceptance model. The effectiveness and the use of outcomes have an impact on the daily routine. This suggests that technology users will consider it as a part of their daily routine if it improves their extrinsic motivation. Additionally, it can support the integration of technology with work processes [13]. However, the impact of daily routine varies for different users, because they may have distinct work situations and views about the integration of technology in their routine work [54]. The following hypothesis was hence developed:

**H4:** PR has a positive and significant impact on POA.

### 3.4 Self-Efficiency (SE)

Albert Bandura was the first person to present the idea of self-efficiency, suggesting that the concept was part of the social cognitive theory and was a vital prerequisite for effective learning behavior. Self-efficiency examines the users' perception regarding their ability to perform various tasks and complete these tasks appropriately [55], [56]. There is a close relationship between e-learning systems and the users' self-efficiency in the classroom. The teachers' capability to efficiently and effectively use technology determines the actual teaching practices in classrooms. Therefore, if teachers have high efficiency, they will be able to perform the desired tasks accurately. This goal can be attained by involving students in different activities, and this will serve as an encouragement for teachers to use technologies more consistently and to slowly enhance their proficiency [57]–[60]. It has been observed during the COVID-19 pandemic that self-efficiency in the educational setting has been influenced by the pandemic situation. It was suggested by various researchers that COVID-19 has had an impact on self-efficiency. According to [61], during the pandemic crisis, work commitment among teachers has been influenced by self-efficiency. It was also stated by [62] that self-efficiency potentially impacts technology adoption during COVID-19. The authors asserted that the adoption can be curtailed or even avoided when there is awareness regarding the drawbacks of the pandemic. Therefore, self-efficiency may play a protective role during the pandemic, because it may give rise to a more flexible environment that supports technology adoption. It seems that the vaccination is a solution to prevent the pandemic spread; however, several people refuse to take it because of various reasons. One of the main reasons is the lack of trust in the health system [63]–[65]. Considering the fact that the fear of vaccination affects the mental and physical health of the users, the following was hypothesised:

**H5:** SE has a positive and significant impact on POA.

### 3.5 Flow Theory

Csikszentmihalyi presented the flow theory as a way of understanding the motivation of users. There is a close relationship between motivation and the psychological state in which the cognitive feeling of motivation and efficiency manages the users [66], [67]. The flow theory signifies the state in which users are excessively involved in a given activity. Users gain a highly enjoyable experience when using the technology; therefore, they will do it in any situation. There is a relationship between flow theory and intrinsic motivation, and particularly self-motivation. It is believed that self-motivation is one of the best ways in which users can be encouraged to do various activities while achieving inner satisfaction. Current research suggests that positive attitudes and high effectiveness is attained from flow. Further, it supports high education objectives achievements since students become motivated to attain specific objectives [68]–[70]. Since the COVID-19 pandemic occurred, various researchers have presented similar results. The motivation of the students during the pandemic was found to affect and be indicative of their satisfaction, success and learning outcomes [71], [72]. User performance is influenced by the fear of vaccination that is expected to reduce the pandemic's bad effects. Hence, the following was hypothesised:

**H6:** EJ has a positive and significant impact on POA.

### 3.6 Critical Mass Theory (PC)

According to the critical mass theory, a significant contribution is made by a population group when specific actions are being adopted. Hence, the behaviour is considered essential by the other individual and the same behaviour is imitated. The critical mass influence for technology adoption is quite vital. If a group of friends has decided to make use of a technology, it is expected that the next group would also adopt it [73]–[75]. Hence, the following hypothesis was formed:

**H7:** PC has a positive and significant impact on POA.

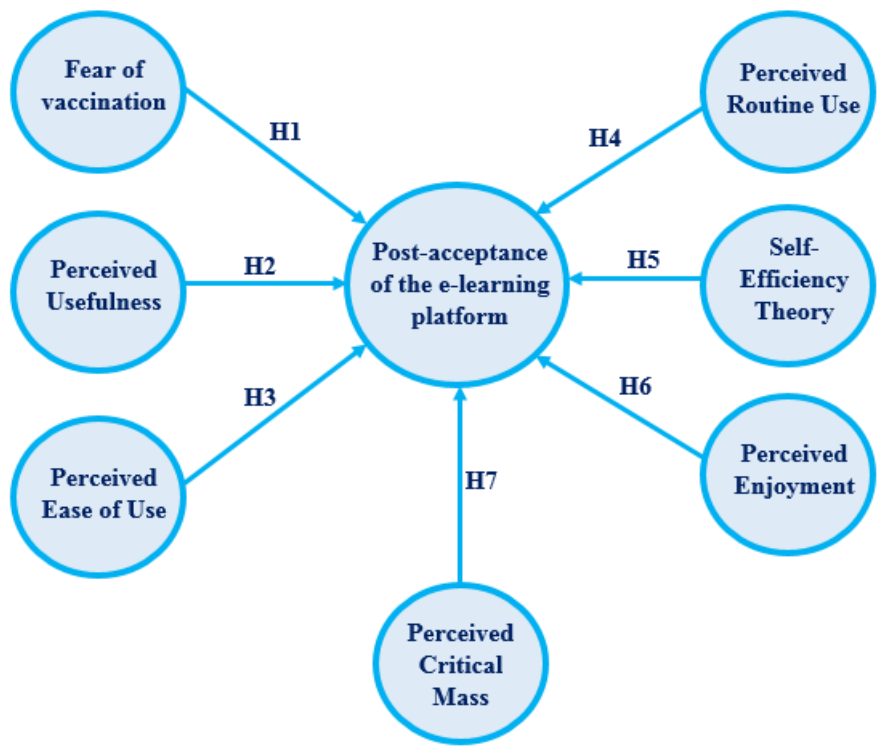

**Figure 1.** The conceptual framework.

## 4. Methods

### 4.1 Statistical Population and Data Collection

The students from UAE universities were given online surveys to fill. The data was gathered during the winter semester 2020/2021, from 25 January 2021 to 30 May 2021. 700 questionnaires were randomly distributed by the research team, out of which 659 questionnaires were filled by the respondents. Hence, the response rate stood at 94%. Since there were a few missing values, 44 of the filled questionnaires were rejected. Therefore, 659 questionnaires were considered to be useful and effective by the team. The sample size of 659 valid responses is quite appropriate as stated by

[76]. The projected sampling size for the 1500 population as part of 306 respondents. When compared to the insignificant requests, the sample size of 659 is quite high. Hence, as the sample size, the structural equation modeling would be accepted [77], and this was needed for hypotheses' confirmation. Additionally, the current theories have been used to establish the hypotheses; yet, they have been included as part of the M-learning context. The measurement model was evaluated using SEM (SmartPLS Version 3.2.7). Advanced treatment was carried out by applying the final path model.

### 4.2 Demographic Characteristics Information

The respondents' personal/demographic data is presented in Table 1. Among the respondents, 48% were males and 52% females, with age group ranging between 18 and 29 years. Nearly 34% of the respondents were older than 29 years. Most respondents had a university degree and a sound educational background. 62% held a bachelor's degree, 23% a master's degree, 12% a doctoral degree and the rest held diplomas. The voluntary respondents of the research [78] recommended that a "purposive sampling approach" be used. The study sample was developed using respondents of various ages and from different colleges, programmes or levels. The IBM SPSS Statistics ver. 23 was used to measure the respondents' demographic information.

**Table 1.** The profile of respondents.

| Criterion | Factor | Frequency | Percentage |
|---|---|---|---|
| Gender | Female | 342 | 52% |
|  | Male | 317 | 48% |
| Age | 18-29 | 436 | 66% |
|  | 30-39 | 97 | 15% |
|  | 40-49 | 94 | 14% |
|  | 50-59 | 32 | 5% |
| Academic qualification | Diploma | 20 | 3% |
|  | Bachelor | 406 | 62% |
|  | Master | 155 | 23% |
|  | Doctorate | 78 | 12% |

### 4.3 Research Instrument

The survey is the instrument which was developed to test the hypotheses of the research. There are 23 items part of the survey which assessed eight constructs present within the questionnaire. Table 2 indicates the sources of all eight constructs. Earlier studies' research questions, after being subjected to some modifications, have been included to ensure the research is applicable.

Table 2. Measurement Items.

| Constructs | Sources |
|---|---|
| POA | [79] |
| PE | [80], [81] |
| PU | [80], [81] |
| PR | [13] |
| EJ | [18], [19] |
| PC | [14], [82] |
| SE | [15], [16] |
| FV | [83], [84] |

## 4.4 Pre-testing the questionnaire

A pilot study was carried out to check the reliability of the questionnaire items. Accordingly, 70 students were randomly selected from the mentioned population. Considering the total sample size to be 10% for the current analysis, the student sample size of 700 was decided, and the research standards were considered appropriately. To assess the pilot study's findings, the internal reliability was tested using Cronbach's alpha (CA) test along with the IBM SPSS Statistics ver. 23. Hence, for the measurement items, acceptable conclusions are presented. It is acceptable to have a 0.70 reliability coefficient if the social science research studies mentioned pattern is emphasised upon [79]. Table 3 presents the seven measurement scales Cronbach alpha values.

Table 3. Cronbach's alpha test.

| Items | CA (≥ 0.70) |
|---|---|
| POA | 0.822 |
| PE | 0.860 |
| PU | 0.724 |
| PR | 0.867 |
| EJ | 0.802 |
| PC | 0.770 |
| SE | 0.866 |
| FV | 0.833 |

## 4.5 Survey Structure

The questionnaire survey was circulated by the researcher. Online surveys were received by the students within United Arab Emirates (UAE) universities (N = 700). Two kinds of popular UAE universities were approached within this research.

The questionnaire survey given to the students [78] had three sections.

- The first section emphasises upon the respondent's personal data.
- The second section included the two items which indicate the general questions regarding e-learning systems.
- The third section included 21 items related to "the fear of vaccination, perceived ease of use, perceived usefulness, perceived routine use, perceived enjoyment, perceived critical mass and self-efficiency".

A five-point Likert Scale was used to measure the 23 items with the following indicators: strongly disagree (1), disagree (2), neutral (3), agree (4) and strongly agree (5).

## 5. Findings and Discussion

### 5.1 Data Analysis

The earlier empirical studies used the SEM to carry out a single-stage analysis. For the current research, a hybrid SEM-ANN approach using deep learning for validation of the hypothesised associations between the research model factors was applied. There are two phases within this research. At first, the recommended research model would use the partial least squares-structural equation modeling (PLS-SEM) by applying SmartPLS [80]. The objective of using PLS-SEM within the research is the theoretical model exploratory nature and the non-availability of earlier literature. General guidelines were followed by the study for implementing the PLS-SEM within the information systems research [81]. It has already been mentioned [82] that the research model can be analysed using a two-step approach (i.e., measurement model and structural model).

Further, IPMA was applied within this research as it is an advanced PLS-SEM technique that helps extract the performance and importance of each construct within the research model. Second, the PLS-SEM analysis investigation, complement, and authentication would be done through ANN. It would also help state the independent variables effectiveness upon the dependent variable. The ANN is considered to be an instrument for function approximation, which is relevant in areas where the collaboration between input(s) and output(s) is non-linear as well as complex. ANN states that there are three vital mechanisms: network architecture, learning rule and transfer function [82]. It is then divided into further subcategories such as "radian basis, feed-forward multilayer perceptron (MLP) network and recurrent network" [32]. One commonly used approach is the MLP neural network in which several layers are present, such as input and output. Hidden nodes are used to connect these input and output layers. Neurons or independent variables are present within the input layer, and they are responsible for taking the raw data ahead to the hidden layers as synaptic weights. The activation function choice determines the hidden layer output. A widely used activation function is the sigmoidal function [83], [84]. The MLP neural network has been used for the recommended research model training and testing.

## 5.2 Convergent Validity

The measurement model was assessed keeping in mind the construct reliability (which integrates "composite reliability (CR), Dijkstra-Henseler's (PA), and Cronbach's alpha (CA)") along with validity (which integrates convergent and discriminant validity). This recommendation was brought forward by [85]. The construct reliability was determined using Table 4 which presents the CA values of 0.769 till 0.906. As compared to the threshold value of 0.7 [86], these figures are higher. Table 4 indicates that the CR values are present between 0.786 and 0.927, and these are also greater than the threshold or recommended value of 0.7 [87]. The Dijkstra-Henseler's rho (pA) reliability coefficient should be used by the research for construct reliability evaluation and reporting.

The reliability coefficient ρA should be similar to the CA and CR with values of 0.70 or higher as part of the exploratory research, and for advanced research stages, it should be over 0.80 or 0.90 [86], [88], [89]. It has been indicated in Table 4 that for each measurement construct, the reliability coefficient ρA should be over 0.70. As the results suggest, the construct reliability should be confirmed and the constructs must be assumed to be quite error-free towards the end.

The convergent validity measurement testing should be done for the average variance extracted (AVE) and factor loading [85]. The findings in Table 4 indicate that the suggested value of 0.7 remained lower than the factor loading values. Further, the AVE-produced values lay between 0.686 and 0.842, and these are much higher than the threshold value of 0.5. For all constructs, it is possible to appropriately attain the convergent validity keeping the future results in mind.

### 5.3 Discriminant Validity

For the measurement of discriminant, it has been recommended that "the Fornell-Larker criterion and the Heterotrait-Monotrait ratio (HTMT)" be measured [85]. The outcomes in Table 5 indicate that the Fornell-Larker condition confirms the requirements as the AVEs and their square roots are much higher than the correlation with the rest of the constructs [90]. Table 6 illustrates the HTMT ratio results, which indicate the fact that 0.85 is the threshold value which is ahead of each construct's value [91]. Therefore, the HTMT ratio is created. Based on these findings, the discriminant validity is stated. Keeping in mind the analysis outcomes, there were no problems related to the measurement model's assessment in terms of the reliability and validity. Hence, it is possible to assess the structural model by making further use of the collected data.

Table 4. Assessment of the measurement model.

| Constructs | Items | Factor Loading | Cronbach's Alpha | CR | PA | AVE |
|---|---|---|---|---|---|---|
| Post-Acceptance of E-learning Technology | POA1 | 0.815 | 0.786 | 0.793 | 0.864 | 0.760 |
| | POA2 | 0.876 | | | | |
| Perceived Routine Use | PR1 | 0.691 | 0.788 | 0.790 | 0.877 | 0.706 |
| | PR2 | 0.876 | | | | |
| | PR3 | 0.881 | | | | |
| Perceived Ease of Use | PE1 | 0.742 | 0.867 | 0.877 | 0.919 | 0.790 |
| | PE2 | 0.873 | | | | |
| | PE3 | 0.863 | | | | |
| Perceived Usefulness | PU1 | 0.841 | 0.892 | 0.927 | 0.931 | 0.819 |
| | PU2 | 0.825 | | | | |
| | PU3 | 0.843 | | | | |
| Perceived Enjoyment | EJ1 | 0.904 | 0.906 | 0.911 | 0.941 | 0.842 |
| | EJ2 | 0.914 | | | | |
| | EJ3 | 0.934 | | | | |

| | | | | | | | |
|---|---|---|---|---|---|---|---|
| Perceived Critical Mass | PC1 | 0.889 | 0.880 | | 0.890 | 0.926 | 0.807 |
| | PC2 | 0.854 | | | | | |
| | PC3 | 0.728 | | | | | |
| Self –efficiency | SE1 | 0.924 | 0.769 | | 0.786 | 0.867 | 0.686 |
| | SE2 | 0.867 | | | | | |
| | SE3 | 0.902 | | | | | |
| Fear of Vaccination | FV1 | 0.867 | 0.844 | | 0.847 | 0.906 | 0.763 |
| | FV2 | 0.896 | | | | | |
| | FV3 | 0.856 | | | | | |

Table 5. Fornell-Larcker Scale.

| | POA | PE | PU | PR | EJ | PC | SE | FV |
|---|---|---|---|---|---|---|---|---|
| POA | **0.873** | | | | | | | |
| PE | 0.611 | **0.898** | | | | | | |
| PU | 0.731 | 0.552 | **0.889** | | | | | |
| PR | 0.714 | 0.482 | 0.699 | **0.918** | | | | |
| EJ | 0.694 | 0.685 | 0.689 | 0.708 | **0.840** | | | |
| PC | 0.721 | 0.703 | 0.724 | 0.594 | 0.646 | **0.905** | | |
| SE | 0.678 | 0.450 | 0.753 | 0.586 | 0.606 | 0.647 | **0.872** | |
| FV | 0.446 | 0.693 | 0.735 | 0.671 | 0.717 | 0.735 | 0.617 | **0.828** |

Table 6. Heterotrait-Monotrait Ratio (HTMT).

| | POA | PE | PU | PR | EJ | PC | SE | FV |
|---|---|---|---|---|---|---|---|---|
| POA | | | | | | | | |
| PE | 0.706 | | | | | | | |
| PU | 0.653 | 0.627 | | | | | | |
| PR | 0.614 | 0.533 | 0.686 | | | | | |
| EJ | 0.652 | 0.620 | 0.631 | 0.539 | | | | |
| PC | 0.514 | 0.505 | 0.697 | 0.637 | 0.407 | | | |
| SE | 0.689 | 0.573 | 0.668 | 0.643 | 0.619 | 0.701 | | |
| FV | 0.445 | 0.672 | 0.696 | 0.398 | 0.534 | 0.585 | 0.642 | |

## 5.4 Fitness of Structural Model

The fit measures offered by SmartPLS are "standard root mean square residual (SRMR), exact fit criteria, d_ULS, d_G, chi-square, NFI, and RMS_theta" which shows the PLS-SEM model fit [92]. Through the SRMR, it is possible to indicate the difference amongst the observed correlations and model implied correlation matrix [93]. The good model fit measures are the values below 0.08 [94]. A good model fit is when the NFI values are greater than 0.90 [95]. The chi-square value ratio is the NFI for the recommended model to the benchmark or null model [96]. The NFI would be highly based on the large parameters, which is why it is not considered to be an appropriate model fit indicator [93]. The discrepancy amongst empirical covariance matrix and the covariance matrix indirect to the composite factor model can be extracted through the two metrics: the squared Eucledian distance d_ULS and the geodesic distance d_G [93], [97]. The RMS theta can only be applied to the reflective models which assess the outer model residuals' correlation degree [96]. The PLS-SEM model is considered efficient if the RMS theta is close to zero. The values are a good fit if they are lower than 0.12; but for any other value, it indicates a lack of fit [98]. The correlation amongst the constructs can be evaluated through the saturated model according to [93]; however, the model structure and total effects are only observed by the estimated model.

According to Table 7, 0.070 is the RMS_theta value, which indicates that the global PLS model validity can be demonstrated due to the large size of the related goodness-of-fit for PLS-SEM model.

Table 7. Measurement model fit indices.

|  | Complete Model | |
| --- | --- | --- |
|  | Saturated Model | Estimated Mod |
| SRMR | 0.035 | 0.035 |
| d_ULS | 0.771 | 1.249 |
| d_G | 0.530 | 0.530 |
| Chi-Square | 473.629 | 473.629 |
| NFI | 0.851 | 0.851 |
| Rms Theta | 0.070 | |

## 5.5 Hypotheses Testing using PLS-SEM

The Smart PLS, with maximum likelihood estimation, was integrated with the structural equation model to extract the several structural model theoretical constructs interdependence [99], [100]. Hence, it was possible to assess the proposed model. It is observed in Table 8 as well as Figure 2 that a moderate predictive power is present for the model, since it attains a 45% variance percentage for post-acceptance of e-learning technology.

The beta ($\beta$) values, $t$-values, and $p$-values for each hypothesis, considering the PLS-SEM technique generated results, are stated in Table 9. It can be stated that all hypotheses have been supported by the researchers. Considering the data analysis, the empirical data supports the H1, H2, H3, H4, H5, H6, and H7 hypotheses.

The results revealed that post-acceptance of e-learning technology (POA) significantly influenced fear of vaccination (FV) (β = 0.293, t = 13.694), perceived usefulness (PU) (β = 0.487, t = 3.494), perceived ease of use (PE) (β = 0.597, t = 17.001), perceived routine use (PR) (β = 0.389, t = 7.009), self-efficiency (SE) (β = 0.548, t = 5.416), perceived enjoyment (EJ) (β = 0.363, t = 6.907) and perceived critical mass (PC) (β = 0.287, t = 5.312), supporting hypotheses H1, H2, H3, H4, H5, H6 and H7, respectively.

Table 8. The $R^2$ values

| Constructs | $R^2$ | Results |
|---|---|---|
| POA | 0.450 | Moderate |

Table 9. Hypotheses testing results.

| H | Path | Path | t-value | p-value | Direction | Hyp/supported | Sig. |
|---|---|---|---|---|---|---|---|
| H1 | FV-> POA | 0.293 | 13.694 | 0.000 | Positive | Yes | ** |
| H2 | PU-> POA | 0.487 | 3.494 | 0.036 | Positive | Yes | * |
| H3 | PE-> POA | 0.597 | 17.001 | 0.000 | Positive | Yes | ** |
| H4 | PR-> POA | 0.389 | 7.009 | 0.003 | Positive | Yes | ** |
| H5 | SE-> POA | 0.548 | 5.416 | 0.008 | Positive | Yes | ** |
| H6 | EJ-> POA | 0.363 | 6.907 | 0.007 | Positive | Yes | ** |
| H7 | PC-> POA | 0.287 | 5.312 | 0.000 | Positive | Yes | ** |

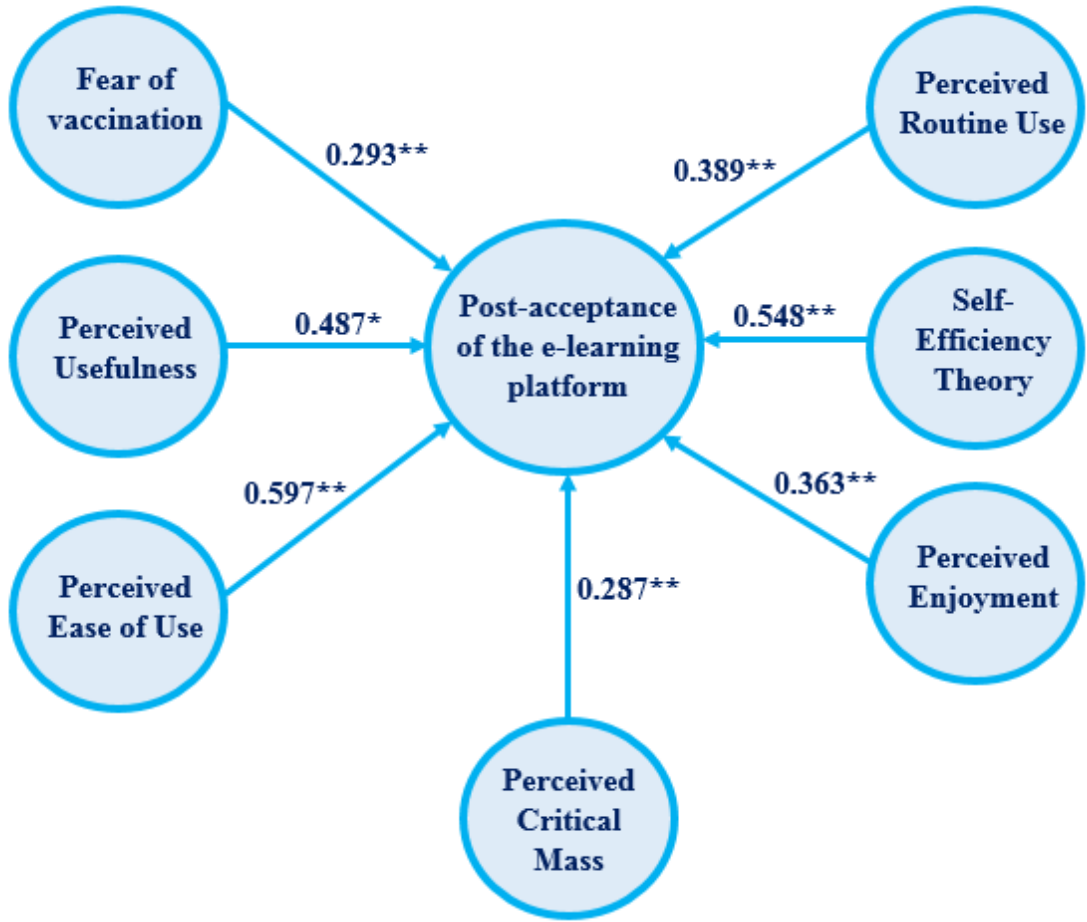

**Figure 2.** Structural model and paths coefficient.

### 5.6 ANN Results

The IBM SPSS Statistics ver. 23 was used to conduct the ANN review. Only the notable predictors were produced from the PLS-SEM findings, which were then used in the ANN evaluation. For the ANN analysis, only the FV, PR, EN, SF, PC, PE, and PU variables were taken into account. The ANN model, as per Fig. 3, has one output neuron (e.g., post-acceptance of e-learning technology) and several input neurons (i.e., FV, PR, EN, SF, PC, PE, and PU). To enable deeper learning and to take effect for each of the output neuron nodes, a two-hidden layer deep ANN model was used [101]. The sigmoid function was used as the activation function for both hidden and output neurons in this study. To improve the efficiency of the presented research model [102], the spectrum for both input and output neurons was specified between [0, 1]. A tenfold cross-validation approach with a ratio of 80:20 for both training and testing data was used to prevent overfitting in ANN models [83]. The root mean square of error (RMSE) was proposed as a measure of the neural network model's accuracy. The RMSE parameters of the ANN model for both training and testing data were 0.1394 and 0.1405, respectively, as seen in Table 4. Since the RMSE estimates and standard deviation for both training and testing data were of minuscule variances(0.0046 and 0.0098, respectively), it can be assumed that the presented research model achieves high accuracy with the use of ANN.

## 5.7 Sensitivity Analysis

The average of every predictor was compared to the maximum mean value expressed as a percentage to determine the normalised importance. Every one of the predictors involved in ANN modelling is included in Table 5 with their mean importance and normalised importance. The sensitivity analysis results show that PE is by far the most important predictor of post-acceptance of e-learning technology, preceded by FV, PR, EN, SF, PC, and PU, as shown in Table 5. It was proposed that the goodness of fit, which is identical to $R^2$ in PLS-SEM analysis [103], be determined to additionally authenticate and validate the ANN application's accuracy and performance. As an outcome, the predictive power of ANN evaluation ($R^2 = 79\%$) was far greater than that of PLS-SEM ($R^2 = 45\%$), according to the findings. These results show that the ANN model elucidates endogenous constructs more thoroughly than the PLS-SEM approach. Further, the disparity in variances can be due to the deep learning ANN approach's dominance in deciding non-linear relationships between the constructs.

**Table 11:** Independent Variable Importance

|    | Importance | Normalized Importance |
|----|------------|-----------------------|
| PE | .270       | 100.0%                |
| FV | .181       | 67.1%                 |
| PR | .163       | 60.4%                 |
| EN | .135       | 49.9%                 |
| SF | .086       | 31.7%                 |
| PC | .085       | 31.5%                 |
| PU | .082       | 30.2%                 |

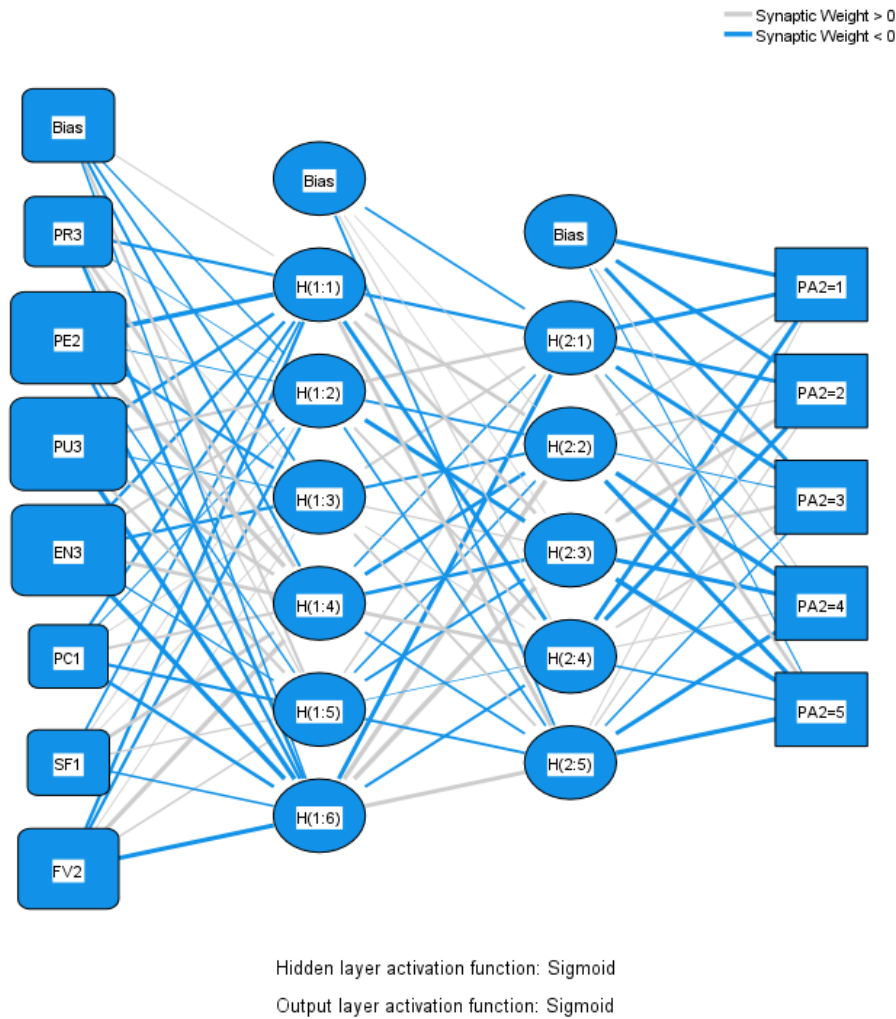

**Figure 3.** ANN model.

### 5.8 Importance-Performance Map Analysis

The IPMA was used as an efficient approach in PLS-SEM with post-acceptance of e-learning technology as the target variable in this study. IPMA, according to Ringle and Sarstedt [104], aids in the comprehension of PLS-SEM research evidence. IPMA includes the average value of the latent constructs and their indicators (e.g., performance measure) as an option to just measure the path coefficients (e.g., importance measure) [104]. According to IPMA, the overall impact represents the importance of precedent factors in defining the target factor (post-acceptance of e-learning technology), while the average of latent constructs' values reflects their performance. The IPMA findings are illustrated in Figure 4. The importance and performance of the seven variables were measured in this study. According to the findings, PE has the maximum values for both importance and performance measures. Notably, FV has the second-largest values in terms of importance and performance measures. Further, although PU has the third-largest value on the importance measure, it has the minimum value on the performance measure. Although SE has the lowest importance measure, it is important to note that on the performance measure, it has the highest relative value to PE.

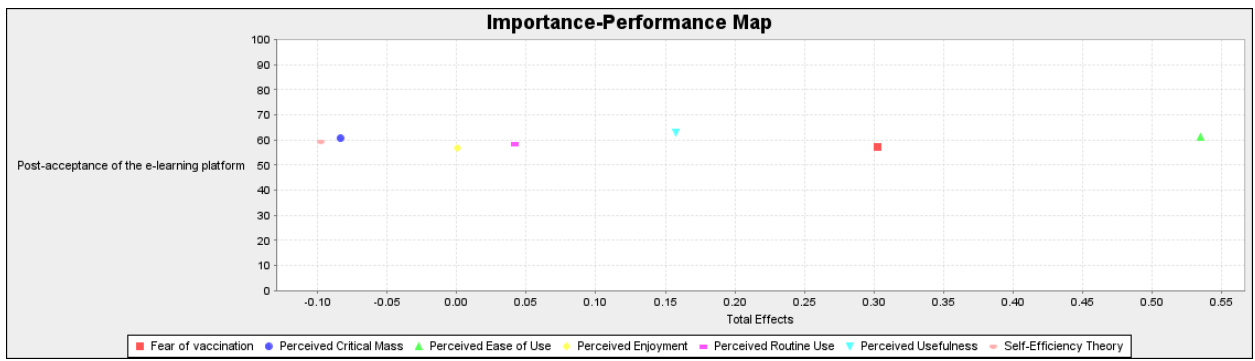

**Figure 4.** IPMA results

## 6. Discussion

The current analysis looked into the key factors that influence the acceptance of an e-learning platform using the TAM model in conjunction with external factors such as "fear of vaccination, perceived ease of use, perceived usefulness, perceived routine use, perceived enjoyment, perceived critical mass and, self-efficiency". The following are some of the most important findings. First, it was discovered that all of the variables have an impact and are linked to the dependent variable. This finding backs up the TAM in the study, which shows that the two constructs of "perceived ease of use and perceived usefulness" have a favourable impact on technology acceptance [23], [105]–[107]. According to a study by [108], perceived ease of use and perceived usefulness have a significant impact on Zoon's acceptance as an e-learning platform, and they have a successful relationship with self-efficacy. The results of this study show that both PEOU and PU have an important and optimistic effect on the e-learning platform's post-acceptance.

Second, the assumptions concerning "perceived routine use and perceived enjoyment" were ensured. These results are consistent with those observed in earlier research. According to [13], the perceived routine use has a beneficial effect on technology acceptance and has a direct association with motivation; similarly, prior research has found that perceived enjoyment has an important impact on technology acceptance [23], [107], [109], [110]. According to an analysis by [111], perceived enjoyment has a favorable impact on technology acceptance, and this factor is closely linked to fast internet and very well systems.

Consequently, the statistical analysis backed up critical mass and self-efficiency. The outcome demonstrates that in moments of distress, people try to be compassionate. They assist one another in avoiding technical difficulties. As a result, these two parameters can act as a buffer throughout a crisis [61], [62], [73], [74]. According to an analysis by [23], self-efficacy and TAM (PEOU and PU) are the most important factors that influence acceptance of technology.

Eventually, according to the results, it was found that fear of vaccination is a major variable that influences the acceptance of an e-learning platform. The results show that fear of vaccination has a beneficial effect on the key model constructs. Vaccine hesitancy has been identified as a critical problem, as it can build numerous obstacles in a post-crisis environment [7]–[9]. Students ought to have vaccination confidence as opposed to vaccination hesitancy; as a result, they would be able to build trust in the healthcare sector, allowing them to tolerate vaccinations and overcome their fears.

## 6.1 Practical Implication

The fear of vaccination is a significant barrier to students' acceptance of technology. As a result, practical implications can aid in the direction of learning techniques and methods. Fear of vaccination, according to experts, may harm users' views. Therefore, education practitioners and teachers should pay close consideration to students' perceptions and preferences while implementing e-learning models [112], [113]. Accordingly, this study has several practical implications. To begin, teachers and healthcare workers should explore various methods for boosting the confidence of technology consumers to accept vaccination. Second, to explore this problem further, investigators may obtain practical confirmation of the impact that vaccination fear can have on the learning process. Further, the current research provides a scientific viewpoint.

Fear of vaccination should be taken into account by academics in educational institutions in the online learning environment, particularly in the teaching and learning surroundings. Accordingly, teachers should rethink the importance of evaluating their teaching strategies to address the current obstacles [114].

### 6.2 Managerial Implication

This study analysed the impact of vaccination hesitancy in the educational environment to encourage healthcare executives and governments to pay attention to vaccination fear and take appropriate measures to alleviate this fear. It provides an in-depth look at existing strategies for reducing fear and boosting vaccination morale among educators, teachers, and students, which will have an impact on society overall. For example, healthcare managers should accept these results to minimise the risk of a crisis that might end in student vaccination rejection [115]. This research has the potential to improve teaching and learning efficiency in educational institutions as well as increase vaccination acceptance, allowing educational institutions to execute their objectives and perspective more efficiently.

### 6.3 Theoretical Implications

In terms of methodology, contrasting prior empirical studies that mainly depended on SEM analysis, this study used a hybrid SEM-ANN model focused on deep learning to refer to the existing research in general and the e-learning domain in particular. The ANN model has a much higher predictive power than the PLS-SEM model. It is concluded that the higher predictive power extracted from ANN analysis results from the deep ANN architecture's power to assess non-linear associations between the factors in the theoretical model.

### 6.4 Limitations of the Study

There are certain drawbacks to the current research. The key drawback is that only four universities in the UAE were engaged, which prevents a thorough examination of the factors influencing e-learning platforms following the propagation of COVID-19. With the involvement of a larger number of universities, the research may have been more useful. By thoroughly analysing the factors that influence e-learning systems, additional analysis would enable an accurate understanding of e-learning systems. Another limitation is the small number of respondents who participated in the study (659 students). According to [78], the survey questionnaire was developed for data collection. The study could have been improved if a superior instrument and sampling method had been used. Further, the participation of various universities from the Arab Gulf region, including those from the KSA, Kuwait, and Bahrain, would have enhanced study results. In the future, it will be imperative to engage more students in participating in research. Further, interviews and focus groups would provide more reliable results. Finally, we hope that the participating Arab universities would adopt an e-learning system.

stability of intention to accept a COVID-19 vaccine in Scotland: Will those most at risk accept a vaccine?," *Vaccines*, vol. 9, no. 1, p. 17, 2021.

[113] K. Cicha, M. Rizun, P. Rutecka, and A. Strzelecki, "COVID-19 and Higher Education: First-Year Students' Expectations toward Distance Learning," *Sustainability*, vol. 13, no. 4, p. 1889, 2021.

[114] V. Cardullo, C. Wang, M. Burton, and J. Dong, "K-12 teachers' remote teaching self-efficacy during the pandemic," *J. Res. Innov. Teach. Learn.*, 2021.

[115] V. Simonetti *et al.*, "Anxiety, sleep disorders and self-efficacy among nurses during COVID-19 pandemic: A cross-sectional study," *J. Clin. Nurs.*, 2021.